\documentclass[%
reprint,
amsmath,amssymb,
aps,prl]{revtex4-2}
\usepackage[latin9]{inputenc}
\usepackage{amsmath}
\usepackage{amssymb}

\usepackage{graphicx}

\usepackage{hyperref}

\begin{document}
\title{Gravitational waves from the instability of cosmological phase transition fronts}

\author{Ariel Mégevand}
\email{megevand@mdp.edu.ar}

\author{Federico Agustín Membiela}
\email{membiela@mdp.edu.ar}

\affiliation{Instituto de Investigaciones Físicas de Mar del Plata (IFIMAR) CONICET,
UNMDP, Funes 3350, 7600 Mar del Plata, Argentina}

\begin{abstract}
We calculate the spectrum of gravitational waves resulting from a first-order electroweak phase transition,
taking into account the hydrodynamic instability of deflagrations. 
This instability causes the walls of bubbles to corrugate and fluid perturbations to grow within a thin shell next to them.
The corrugation length scale is several orders of magnitude smaller than the typical bubble radius.
The result is a peak and a ``knee'' in the gravitational wave spectrum at a frequency several orders of magnitude higher than the bubble collision peak. 
This signal may be detectable by future gravitational wave observatories such as BBO or DECIGO.
\end{abstract}

\maketitle

First-order phase transitions (FOPTs) in the early universe 
proceed through the nucleation and expansion of bubbles in the primordial plasma
\cite{quiros07field,abf+24cosmological}. 
The motion of bubble walls and the fluid generate
a stochastic background of gravitational waves (GWs)
that may be detectable by future GW observatories \cite{hogan86gravitational,tw90relic}.
In particular,
space-based interferometers such as LISA \cite{a+17laser}, BBO \cite{cc06detecting} and DECIGO \cite{k+11japanese} may be able to
detect GWs from a first-order electroweak phase transition \cite{kkt94gravitational,amn+01supersymmetric}.
While this transition is a smooth crossover in
the Standard Model of particle physics \cite{klr+96is}, many extensions
of the Standard Model allow a FOPT (see, e.g., \cite{cqw96opening,gsw05first,ekn+08some,hk08production,dhk+17second,mnq18cosmological}).

There are three well-known mechanisms of GW production during a FOPT,
namely, bubble collisions \cite{ktw92gravitational,kt93gravitational},
turbulence \cite{kkt94gravitational,dgn02relic,kmk02gravitational,cd06gravitational},
and sound waves \cite{hhr+14gravitational,hindmarsh18sound}. For
these mechanisms, the frequency of the GW signal is determined by the duration
of the phase transition, $\omega\sim t_{b}^{-1}$. 
A very different
scale is introduced by a hydrodynamic instability in the propagation
of subsonic bubble walls 
\cite{LandauLifshitz87Fluid,fa90hadron,link92deflagration,kf92instability,hkl+93hydrodynamic,mm14stability},
which causes wall deformations on a length scale much shorter
than the bubble radius, potentially modifying the GW spectrum at frequencies
$\omega\gg t_{b}^{-1}$. 

In this work, we present the first estimate of the GW signal resulting from this instability.
We will consider the electroweak
phase transition.
In Ref.~\cite{Largo} we consider phase transitions at various temperature scales.

In the early universe, the pressure $p$ is related to the temperature
$T$ by the equation of state (EOS). During a FOPT, there are two
coexisting phases, which can be modeled by the bag EOS:
$p_{+}=\frac{\pi^{2}}{90}g_{*}T^{4}-\epsilon$ and
$p_{-}=(\frac{\pi^{2}}{90}g_{*}-\frac{\epsilon}{T_{c}^{4}})T^{4}$.
The index $+$ denotes the phase that is stable at higher temperatures
and the index $-$ denotes the phase that is stable at lower temperatures;
$g_{*}$ is the number of relativistic degrees of freedom of the plasma,
$\epsilon$ is the false vacuum energy density, and $T_{c}$ is the
critical temperature. For the electroweak phase transition, we have
$T_{c}\sim100\,\mathrm{GeV}$ and $g_{*}\sim100$. We will consider
different values of the ratio $\alpha_{T}=\epsilon/\rho_R$,
where $\rho_{R}=\frac{\pi^{2}}{30}g_{*}T^{4}$ is the radiation energy
density.

Below the critical temperature, the metastable phase ($+$) becomes
supercooled, and bubbles of the stable phase ($-$) nucleate and expand. For
simplicity, we will assume that all bubbles nucleate simultaneously.
In Ref.~\cite{Largo}, we consider 
a nucleation rate of the form $\Gamma \propto e^{\beta t}$.
We denote the number density of bubbles as $n_{b}$ and define the parameter
$R_{b}=(4\pi n_{b}/3)^{-1/3}$, which is associated with the average bubble separation.
The wall motion is driven by the pressure difference between phases.
Due to friction with the plasma, the wall quickly reaches a terminal velocity $v_w$
which depends on the model. 
We will treat it as a free parameter.
At time $t$ since nucleation, the bubbles have a radius $R=v_{w}t$. 
The fraction of the volume occupied by the $+$ phase is given by 
\begin{equation}
f_{+}=e^{-I},\quad I={\textstyle \frac{4\pi}{3}}n_{b}v_{w}^{3}t^{3}=t^{3}/t_{b}^{3},
\end{equation}
where $t_{b}=R_{b}/v_{w}$ is the timescale of the phase transition. 
The parameter $t_{b}$ can be linked to the parameter $\beta$ of
the exponential nucleation through a physical definition
of phase transition duration \citep{mm21model,Largo}.
We have $\beta t_b\simeq 4$ \cite{mm21model,Largo}.
Typically, $\beta\sim 10-1000H$, where 
$H=(\frac{8\pi}{3}G\rho_{\mathrm{tot}})^{1/2}$ is the Hubble rate,
$G$ is Newton's constant and $\rho_{\mathrm{tot}}=\rho_{R}+\epsilon$
is the total energy density.

For subsonic velocities ($v_{w}<1/\sqrt{3}$), the wall propagates
as a deflagration, which is known to be unstable \cite{LandauLifshitz87Fluid,fa90hadron,link92deflagration,kf92instability,hkl+93hydrodynamic,mm14stability}.
Stability of stationary propagation has rarely been
studied beyond planar walls.
However, we can apply the results obtained for planar walls
since the length scale of unstable perturbations is much smaller than the bubble radius.
Wall deformations are described by $z=\zeta(\mathbf{x},t)$, where
$\mathbf{x}$ denotes a point on the unperturbed wall.
The equation for $\zeta$ must be solved alongside those
for the fluid perturbations in each phase,
using appropriate boundary and matching conditions. 
Assuming solutions of the form $e^{\Omega t+i\mathbf{\mathbf{k}\cdot\mathbf{x}}}$
for $\zeta$ and $A_{i}e^{\Omega t+i\mathbf{\mathbf{k}\cdot\mathbf{x}}+q_{i}z}$
for the various fluid variables, we obtain a system of algebraic equations
for the coefficients. 
There are two physical solutions, with $\Omega$ given by \cite{mm14stability}
\begin{equation}
\Omega_{-}\simeq-(v_{w}d)^{-1},\quad\Omega_{+}\simeq {v_{w}k_{c}}d(1-{k}/{k_{c}})k \label{Ommas}
\end{equation}
(for simplicity, we use the approximations $v_{w}\ll1$ and $\alpha_{T}\ll1$).
Here, $k=|\mathbf{k}|$, 
$k_{c}\simeq\frac{3}{2}\alpha_{T} (1-{v_{w}^{2}}/{v_{c}^{2}})d^{-1}$,
and $d=\sigma_{0}/(4\epsilon v_{c}^{2})$,
where 
$\sigma_{0}$ is the surface tension
and
$v_c$ is a critical velocity.
Typically, we have $\sigma_{0}\sim 0.1 T_c^{3}$ for the electroweak phase transition.
The critical velocity $v_c$ has a different parametric dependence than $v_w$,
and we will treat it as an independent parameter (see details in \cite{Largo}).
The quantity $\Omega_{-}$ is always negative. 
On the other hand,
we have $\Omega_{+}>0$ for $v_{w}<v_{c}$ and $k<k_{c}$. Therefore,
below the critical velocity $v_{c}$, perturbations of wavelength $k^{-1}>k_{c}^{-1}$
grow exponentially. We have $k_{c}^{-1}\sim T^{-1}\ll R_{b}$.

In the presence of time-dependent sources of perturbation, we may assume wall deformations of the form
$\zeta=D_{\mathbf{k}}(t)e^{i\mathbf{k}\cdot\mathbf{x}}$
and similar forms for the fluid perturbations.
A general deformation is given by the Fourier integral
\begin{equation}
	\zeta (\mathbf{x},t) = \int \negthickspace \frac{d^{2}k}{(2\pi)^{2}}D_{\mathbf{k}}(t)e^{i\mathbf{k}\cdot\mathbf{x}} 
	\label{eq:zTFD}
\end{equation} 
(with $D_{-\mathbf{k}}=D_{\mathbf{k}}^{*}$).
The equation for $D_{\mathbf{k}}(t)$ is of the form
\begin{equation}
\sigma_{0}\ddot{D}_{\mathbf{k}}+\eta_{k}\dot{D}_{\mathbf{k}}+\sigma_{0}\omega_{k}^{2}D_{\mathbf{k}}=y_{\mathbf{k}}.\label{oscarm}
\end{equation}
The homogeneous equation has solutions $e^{\Omega_{\pm}t}$ with 
$\Omega_{\pm}=-\eta_{k}/2\sigma_{0}\pm ({\eta_{k}^{2}/4\sigma_{0}^{2}-\omega_{k}^{2}})^{1/2}$.
Inverting this relation yields the coefficients $\omega_{k}$ and $\eta_{k}$
in terms of $\Omega_{+}$ and $\Omega_{-}$, which are given by Eq.~(\ref{Ommas}) (see \cite{Largo} for details).
We	 consider a stochastic
source $y_{\mathbf{k}}$ corresponding to thermal fluctuations, with the properties $\langle y_{\mathbf{k}}\rangle =0$ and
\begin{equation}
	\langle y_{\mathbf{k}}(t)y_{\mathbf{k}'}^{*}(t')\rangle=2\eta_{k}T\delta(t-t')(2\pi)^{2}\delta(\mathbf{k}-\mathbf{k}'),\label{propsfza}
\end{equation}
where $\langle\,\rangle$ denotes ensemble average. 
The specifics of the initial fluctuations are unimportant
since a perturbation only becomes relevant once it reaches the nonlinear regime.
The differential equation (\ref{oscarm}) is that of a
driven harmonic oscillator. 
Given the initial conditions $D_{\mathbf{k}}(t_{0})=\dot{D}_{\mathbf{k}}(t_{0})=0$,
the solution is
\begin{equation}
D_{\mathbf{k}}(t) = 
\int_{t_{0}}^{t} 
\frac{e^{\Omega_{+}(t-\tilde{t})}-e^{\Omega_{-}(t-\tilde{t})}}{\sigma_{0}(\Omega_{+}-\Omega_{-})}y_{\mathbf{k}}(\tilde{t})d\tilde{t}
.\label{solDk}
\end{equation}

We assume that mode $k$ becomes active once its growth rate
exceeds the bubble's expansion rate \cite{soj97primordial}, i.e.,
$\Omega_{+}\geq v_{w}/R$. 
Thus, the activation time
is given by $t_{0}=\Omega_{+}(k)^{-1}$. 
At a given time $t$, the active modes correspond to the range $k_{\inf}<k<k_{\sup}$, with $k_{\sup}\simeq k_{c}$
and $k_{\inf}\sim R^{-1}$.

From Eqs.~(\ref{propsfza}) and (\ref{solDk}), we obtain
\begin{align}
\langle D_{k}(t)D_{k'}^{*}(t')\rangle &=(2\pi)^{2}\delta(\mathbf{k}-\mathbf{k}')|D|^{2}(k,t,t'),\label{defD2}
\\
|D|^{2}(k,t,t') &=D_{0}^2 e^{\Omega_{+}(t-t_{0})} e^{\Omega_{+}(t'-t_{0})},
\end{align}
with 
$D_{0}^{2}\simeq {5\pi^{-2}g_{*}^{-1}\alpha_{T}^{-2}v_{c}^{-2}}{(1-v_{w}^{2}/v_{c}^{2})^{-1}T^{-3}k^{-1}}$.
To obtain this result, we neglected terms containing the factors $e^{\Omega_{-}(t-\tilde{t})}$ and
$e^{-2\Omega_{+}(t-t_{0})}$, which decay rapidly.
We also took into account that $\Omega_{+}\ll|\Omega_{-}|$ and
that the relevant wavenumbers satisfy $k\ll k_c$ (see below). 

The approximation of small perturbations is valid when $\partial_{i}\zeta\ll1$.
For a superposition of modes, we consider the displacement spectrum
$\zeta_{k}$ defined by $\langle\zeta^{2}\rangle\equiv\int\zeta_{k}^{2}\,{\rm d}\ln k$,
and impose the condition $k\zeta_{k}\ll1$ \cite{soj97primordial},
which implies $|D|(k,t,t)\ll k^{-2}$. 
Since the nonlinear regime has not been studied,
we assume that this quantity stops growing when
$|D|(k,t,t)=a k^{-2}$, with $a\lesssim 1$. 
This occurs at the time $t_{nl}(k)$ such that $D_{0}e^{\Omega_{+}(t_{nl}-t_{0})}=ak^{-2}$.
We discuss this approximation in Ref.~\cite{Largo}.
Thus, we have
\begin{equation}
	|D|^{2}(k,t,t')=\frac{a^{2}}{k^{4}}\begin{cases}
		e^{\Omega_{+}(t+t'-2t_{nl})} & t<t'<t_{nl}(k),\\
		e^{\Omega_{+}(t-t_{nl})} & t<t_{nl}(k)<t',\\
		1 & t_{nl}(k)<t<t'.
	\end{cases}\label{D2final}
\end{equation}
For fixed $t$ and $t'$, $|D|^2$ grows exponentially with $k$ for small
$k$ (since we have $\Omega_{+}\propto k$) and decreases as
$k^{-4}$ for large $k$. 
The peak occurs at $k=k_{nl}(t)$, where
$k_{nl}(t)$ is the solution of the equation $t_{nl}(k)=t$.
We have \cite{Largo} $k_{nl}\simeq k_{*}\,t_{b}/t$,
where $k_{*}=k_{nl}|_{t=t_{b}}$. The constant $k_*$ gives the relevant
scale of unstable wall corrugations.
In general, we have $R_b^{-1}\ll k_* \ll k_c$.
Roughly, $k_* \sim 10 \alpha_T^{-1} R_b^{-1}$.

We denote $ \zeta(\mathbf{x},t)$ as $\zeta$ and $\zeta(\mathbf{x}^{\prime},t')$ as $\zeta'$.
The spacetime correlation function $\langle\zeta\zeta'\rangle$ can be obtained from Eqs.~(\ref{eq:zTFD}) and (\ref{defD2})
as a function of $t$, $t'$, and $x = |\mathbf{x}-\mathbf{x}^{\prime}|$.
It can be seen that $\langle\zeta\zeta'\rangle$ becomes negligible beyond a distance
$x\sim k_*^{-1}\ll R_{b}$.
For $x=0$ and $t=t'$,
taking into account that $D_{0}^{2}\ll a^{2}/k_{*}^{4}$ and $k_{*} \ll k_{\sup}$ (see \cite{Largo} for details),
we obtain
\begin{equation}
\langle\zeta^{2}\rangle\simeq\alpha^{2}t^{2}, \quad
\alpha^{2}
	=\frac{a^{2}}{4\pi}\frac{1+(k_* R_b k_c d)^{-1}}{(k_{*}t_{b})^{2}} .
\label{zeta2med}
\end{equation}
Similarly, we obtain 
$\langle(\nabla\zeta)^{2}\rangle \simeq ({a^{2}}/{2\pi})\log(k_{c}/k_{*})$,
i.e., the quantity $\langle(\nabla\zeta)^{2}\rangle$ is approximately constant.

For a spherical bubble, the deformed surface can be described
by $r=R+\zeta(\hat{r})$, where 
$\zeta$ is the surface displacement at position $R\hat{r}$ relative to the bubble center. 
For GW generation, the relevant part of the bubble's energy-momentum tensor is given by \cite{mm23thin,mm25thin}
\begin{equation}
T_{00} =\sigma\,\delta(r-R-\zeta)\,1_{S}(\hat{r}),
\quad T_{ij}  =T_{00}t_{ij}, 
\label{eq:T00wall}
\end{equation}
where $1_{S}(\hat{r})$ is the indicator function for the uncollided wall surface at time $t$ and
\begin{equation}
t_{ij}  =(\hat{r}_{i}-\partial_{i}\zeta)(\hat{r}_{j}-\partial_{j}\zeta)[1+(\nabla\zeta)^{2}]^{-1}.
\label{tij}
\end{equation}
The surface energy density $\sigma$ increases as vacuum
energy is released within the bubble.
Part of the energy is also used to reheat the plasma and drive bulk fluid motions.
As a consequence, we have $\sigma=\kappa\epsilon R/3$, where the factor $\kappa$ accounts
for the fraction of the released energy that goes to the wall. 
The unstable fluid perturbations decay exponentially from the wall
over a length $q^{-1}\sim k^{-1}$ \cite{mm14stability}. 
Since the relevant modes satisfy $k^{-1}\ll R$, the exponential growth occurs within a thin fluid layer. 
Therefore, we can account for the fluid by including in $\kappa$ the energy fraction that goes to the fluid in the stationary case.
The distribution of energy among bubble collisions and unstable perturbations
 arises from the evolution of the quantity $t_{ij}$.

For a stochastic source in a volume $V$, the energy density of gravitational
waves is given by \cite{mm21gravitational}
\begin{equation}
\frac{d\rho_{GW}}{d\ln\omega} =
\frac{4G\omega^{3}}{\pi} \negmedspace \int_{-\infty}^{\infty}\negthickspace dt \negmedspace \int_{t}^{\infty} \negthickspace dt'
\cos(\omega t_-)\,\Pi(t,t',\omega),\label{rhoGW}
\end{equation}
where $\Pi=V^{-1}\Lambda_{ijkl}(\hat{n})\langle\tilde{T}_{ij}(t,\omega\hat{n})\tilde{T}_{kl}(t',\omega\hat{n})^{*}\rangle$.
Here, $\Lambda_{ijkl}$
is the transverse-traceless projection tensor,
$\hat{n}$ is the direction of observation,
and $\tilde{T}_{ij}$
is the Fourier transform of the total $T_{ij}$ of all the bubbles.
We also use the notation $t_\pm=t'\pm t$. 
For an individual bubble, we have
\begin{equation}
\tilde{T}_{ij}=\sigma R^{2}\negmedspace\int\negmedspace d\hat{r}\,e^{-i\omega\hat{n}\cdot\hat{r}R}e^{-i\omega\hat{n}\cdot\hat{r}\zeta}\,t_{ij}1_{S}(\hat{r}),\label{Tijtil}
\end{equation}
where $d\hat{r}=\sin\theta d\theta d\phi$. We have neglected terms
of order $\zeta/R$ since we have $|\zeta/R|\ll|\partial_{i}\zeta|$.
The quantity $\Pi$ is given by the sum of a single-bubble part
$\Pi_{s}$, where the quantities $\tilde{T}_{ij}$ and $\tilde{T}_{kl}$ belong to the same bubble,
and a two-bubble part $\Pi_{d}$, where they
correspond to different bubbles \cite{jt17gravitational,mm21gravitational}. We have \cite{mm21gravitational,mm21model}
\begin{align}
\Pi_{s}= & \,n_{b}{\textstyle \sigma}R^{2}\sigma'R^{\prime2}\negthickspace\int\negthickspace d\hat{r}\negthickspace
\int\negthickspace d\hat{r}'e^{i\omega\hat{n}\cdot\mathbf{s}}
 	\Lambda_{ijkl}G_{ijkl}P_{u},
 	\label{eq:Pis} 
 \\
\Pi_{d}= & \,n_{b}^{2}{\textstyle \sigma}R^{2}\sigma'R^{\prime2}\negthickspace\int\negthickspace d^{3}l\negthickspace\int\negthickspace d\hat{r}\negthickspace\int\negthickspace d\hat{r}'e^{i\omega\hat{n}\cdot\mathbf{s}}
    \Lambda_{ijkl}G_{ijkl}P_{u},\label{eq:Pid}
\end{align}
with $G_{ijkl}=\langle t_{ij}t_{kl}^{\prime} e^{i\omega\hat{n}\cdot(\zeta'\hat{r}'-\zeta\hat{r})} \rangle$,
$P_{u}=\left\langle 1_{S}(\hat{r})1_{S'}(\hat{r}')\right\rangle $,
$R'=R(t')$, $\zeta'=\zeta(\hat{r}',t')$.
The vector $\mathbf{s}$ connects a point on the
surface $S$ of a bubble at time $t$ and a point on the surface $S'$
of a bubble at time $t'$. 
We have $\mathbf{s}=\mathbf{l}+R'\hat{r}'-R\hat{r}$,
where $\mathbf{l}$ connects the two nucleation centers. 
For $\Pi_{s}$, we have $\mathbf{l}=\mathbf{0}$.

The quantity $P_{u}$ is the joint probability of both points $R\hat{r}$
and $R'\hat{r}'$ remaining uncollided.
It was calculated in Ref.~\cite{mm20bubble}.
For the single-bubble contribution, we have
\begin{equation}
	P_{u}= e^{-J}, \quad 
	J =  t_{b}^{-3}\frac{(t_{+}+t_{s})^{2}(2t_{s}t_{+}-t_{s}^{2}+3t_{-}^{2})}{16 t_{s}},
	\label{Itot}
\end{equation}
where $t_s=s/v_{w}$, with $s=\vert\mathbf{s}\vert$.
In the two-bubble case, $P_u$ contains additional Heaviside
function factors, $\Theta(|\mathbf{l}-R\hat{r}|-R)$ and $\Theta(|\mathbf{l}+R'\hat{r}'|-R)$.

The quantity
$G_{ijkl}$ can be calculated using the expansion 
$t_{ij}=t_{ij}^{0}+t_{ij}^{1}+t_{ij}^{2}+\cdots$, with $t_{ij}^{0}=\hat{r}_{i}\hat{r}_{j}$,
$t_{ij}^{1}=-(\hat{r}_{i}\partial_{j}\zeta+\hat{r}_{j}\partial_{i}\zeta)$,
$t_{ij}^{2}=\partial_{i}\zeta\partial_{j}\zeta-\hat{r}_{i}\hat{r}_{j}(\nabla\zeta)^{2}$.
Thus, we have $G_{ijkl} = \sum G_{ijkl}^{nm}$, 
where 
$G_{ijkl}^{nm}=\langle t_{ij}^{n}t_{kl}^{\prime m} e^{i\omega\hat{n}\cdot(\zeta'\hat{r}'-\zeta\hat{r})} \rangle$.
Inserting in Eqs.~(\ref{eq:Pis})-(\ref{eq:Pid}), we obtain expansions $\Pi_{s,d} = \sum\Pi_{s,d}^{nm}$.
Assuming Gaussian fluctuations, the averages can be expressed in terms
of two-point functions and the quantity $G \equiv \langle e^{i\omega\hat{n}\cdot(\zeta'\hat{r}'-\zeta\hat{r})}\rangle$,
which is given by \cite{mm21gravitational}
\begin{multline}
	G
	=\exp\left\{ -{\textstyle \frac{1}{2}}\omega^{2} \left[(\hat{n}\cdot\hat{r})^{2}\langle\zeta^{2}\rangle
				+(\hat{n}\cdot\hat{r}')^{2}\langle\zeta^{\prime2}\rangle
						\right.\right. \\ \left.\left.
	-2(\hat{n}\cdot\hat{r})(\hat{n}\cdot\hat{r}')\langle\zeta\zeta'\rangle\right]\right\} 
	.\label{eq:G}
\end{multline}

For the lowest-order contributions, $\Pi^{00}_{s,d}$, we have
$\Lambda_{ijkl}G_{ijkl}^{00}=\Lambda G$, where $\Lambda=\Lambda_{ijkl}\hat{r}_{i}\hat{r}_{j}\hat{r}'_{k}\hat{r}'_{l}$.
The last term in the exponent of Eq.~(\ref{eq:G}) vanishes for the two-bubble case. 
In the single-bubble case, this term is non-vanishing only when the point separation
is very small, and we  neglect it. Using Eq.~(\ref{zeta2med}),
we obtain
\begin{equation}
G=\exp\{-{\textstyle \frac12}\alpha^{2}\omega^{2}[(\hat{n}\cdot\hat{r})^{2}t^{2}+(\hat{n}\cdot\hat{r}')^{2}t^{\prime2}]\}.\label{eq:G00}
\end{equation}
For $\alpha=0$, we have $G=1$, which corresponds to the bubble-collision mechanism.
Since the source is isotropic, we can integrate out the direction of observation $\hat n$
in Eqs.~(\ref{eq:Pis})-(\ref{eq:Pid}).
Since $\alpha^2\sim (a^{2}/4\pi)(k_{*}t_{b})^{-2}$,
we have $\alpha^{2}\omega^{2}t^2\ll 1$ even for frequencies
$\omega\gtrsim k_{*}$, so we can assume that $G$ is approximately constant
and evaluate it at $\hat{n}=\hat{s}$ (see \cite{Largo} for details).
This approximation breaks down when $\omega\gg k_{*}$, but
in this case the suppression due to the factor $G$
renders this contribution irrelevant. 
With this approximation, we obtain the same result as for the bubble-collision
mechanism, but with the
additional factor $G$,
\begin{align}
\frac{\Pi_{s}^{00}}{\kappa^2\epsilon^2} & =\frac{\pi v_{w}^{3}}{48t_{b}^{3}}
\int_{t_{-}}^{t_{+}} \frac{dt_s}{t_s^{3}}\sum_{n=0}^{2}\frac{j_{n}(\omega v_{w}t_s)}{(\omega v_{w}t_s)^{n}}P_{n}e^{-J} G, 
\label{Pi00sfinal}
\\
\frac{\Pi_{d}^{00}}{\kappa^2\epsilon^2} & =\frac{\pi v_{w}^{3}}{64t_{b}^{6}}
\int_{t_{-}}^{t_{+}} \frac{dt_s}{t_s^{4}}\frac{j_{2}(\omega v_{w}t_s)}{(\omega v_{w}t_s)^{2}}Qe^{-J} G,
\label{Pi00dfinal}
\end{align}
where $j_{n}$ are the spherical Bessel functions, $P_{n},Q$
are polynomials in the variables $t_s,t_{\pm}$ (see \cite{mm21model}), and
$G=\exp\left[-\frac14\alpha^{2}\omega^{2}(t_s^{2}+t_{+}^{2}t_{-}^{2}/t_s^{2})\right]$.
Finally, we insert these results into Eq.~(\ref{rhoGW}). 
The variable $t_{-}$ can be integrated analytically.
The remaining integrals with respect to $t_{+}$ and $t_s$ must be calculated numerically.

Now, let us consider the higher-order contributions.
There are two order-1 contributions, $G_{ijkl}^{10}$ and $G_{ijkl}^{01}$,
which correspond to the products $t_{ij}^{1}t_{kl}^{\prime0}$ and $t_{ij}^{0}t_{kl}^{\prime1}$,
respectively. However, due to the fact that $\langle\partial_{i}\zeta\rangle$
vanishes, these contributions can be neglected \cite{mm21gravitational}.

There are three second-order contributions. 
Two of them correspond to $t_{ij}^{2}t_{kl}^{\prime0}$ and $t_{ij}^{0}t_{kl}^{\prime2}$, 
which contain terms of the form
$\hat{r}_{k}'\hat{r}_{l}'\partial_{i}\zeta\partial_{j}\zeta$
and $\hat{r}_{i}\hat{r}_{j}\partial_{k}\zeta'\partial_{l}\zeta'$. 
We obtain
\begin{equation}
\Pi^{20}+\Pi^{02}=-{\textstyle \frac{3}{2}} \big( \langle(\nabla\zeta)^{2}\rangle + \langle(\nabla\zeta')^{2}\rangle \big)  \Pi^{00}.
\label{Pi2002}
\end{equation}
As discussed above, the quantity $\langle(\nabla\zeta)^{2}\rangle$
is approximately constant.

The final second-order contribution, $\Pi^{11}$, corresponds to 
$G_{ijkl}^{11}=G\langle t_{ij}^{1}t_{kl}^{\prime1}\rangle$.
We will not consider contributions of higher order. 
The quantity $\langle t_{ij}^{1}t_{kl}^{\prime1}\rangle$ 
contains terms of the form $\hat{r}_{i}\hat{r}_{k}' \langle \partial_{j}\zeta\partial_{l}\zeta' \rangle$,
which vanish in the two-bubble case. 
In the single-bubble case, $\langle\partial_{i}\zeta\partial_{j}\zeta'\rangle$ 
is non negligible only for small angular separations \cite{mm21gravitational}.
Using the approximation $\hat{r}'\simeq\hat{r}$ for the coefficients of these correlation functions,
we obtain
\begin{equation}
	\Lambda_{ijkl}\langle t_{ij}^{1}t_{kl}^{\prime1}\rangle
	= 2 n^2 [\langle\nabla\zeta\cdot\nabla\zeta'\rangle
	-\langle(\mathbf{n}\cdot\nabla\zeta)(\mathbf{n}\cdot\nabla\zeta')\rangle],
	\label{Lttpl}
\end{equation}
where $\mathbf{n}\equiv \hat{n}-(\hat{n}\cdot\hat{r})\hat{r}$, $n^2\equiv \mathbf{n}^2 = 1-c^2$, and $c=\hat{r}\cdot\hat{n}$.
Due to the factors $\langle\partial_{i}\zeta\partial_{j}\zeta'\rangle$ in the integrand in (\ref{eq:Pis}),
we can use the approximation $\hat{r}'\simeq\hat{r}$ in Eq.~(\ref{eq:G}).
At high frequency we have $t'=t$ and, hence, 
$\langle\zeta\zeta'\rangle \simeq \langle\zeta^{2}\rangle \simeq \langle\zeta^{\prime2}\rangle$.
Therefore, instead of the suppression factor (\ref{eq:G00}), we obtain $G\simeq1$ in this case (see \cite{Largo} for details).

In Eq.~(\ref{eq:Pis}), we have $\mathbf{s}=R_{-}\hat{r}+\mathbf{x}$,
where $R_{-}=R'-R$ and $\mathbf{x}=R'(\hat{r}'-\hat{r})$
(see \cite{Largo} for a figure). 
Since  $x\equiv|\mathbf{x}|\ll R,R'$ and $\hat{r}\cdot\mathbf{x}\simeq0$,
we can write
\begin{equation}
\int\negthickspace d\hat{r}'R^{\prime 2}e^{i\omega\hat{n}\cdot\mathbf{x}}
\Lambda_{ijkl}G_{ijkl}^{11}
\simeq 	\negthickspace
\int \negthickspace d^{2}xe^{i\omega\mathbf{n}\cdot\mathbf{x}}
\Lambda_{ijkl}\langle t_{ij}^{1}t_{kl}^{\prime1}\rangle.
\label{Pi11pl}
\end{equation}
We have taken the factor $e^{-J}$ out of this integral since we have
$J\simeq J_{x=0}=J_{s=R_{-}}=t^{\prime3}/t_{b}^{3}$.
The integral in Eq.~(\ref{Pi11pl}) is the Fourier transform of Eq.~(\ref{Lttpl}), 
evaluated at $\mathbf{k}=\omega\mathbf{n}$.
The Fourier transform of $\left\langle \zeta\zeta'\right\rangle $
is given by $|D|^{2}(k,t,t')$, and the derivatives in Eq.~(\ref{Lttpl})
introduce factors $k_{i}k_{j}$. We obtain
\begin{equation}
\frac{\Pi^{11}}{\kappa^2\epsilon^2}=\frac{v_{w}t^{3}t^{\prime}}{3t_{b}^{3}} 
e^{-J} \negthickspace\negthinspace \int_{-1}^{+1}\negthickspace\negthickspace dce^{i\omega v_wt_{-}c}c^{2}(1-c^2)^{2}\omega^{2}|D|^{2}.
\label{Pi11final}
\end{equation} 
Finally, we insert Eq.~(\ref{Pi11final}) into Eq.~(\ref{rhoGW}). 
The variable $t$ can be integrated analytically.
The remaining integrals with respect to $t'$ and $c$ must be calculated numerically. 

\begin{figure}[h]
\centering
\includegraphics[width=0.9\columnwidth]{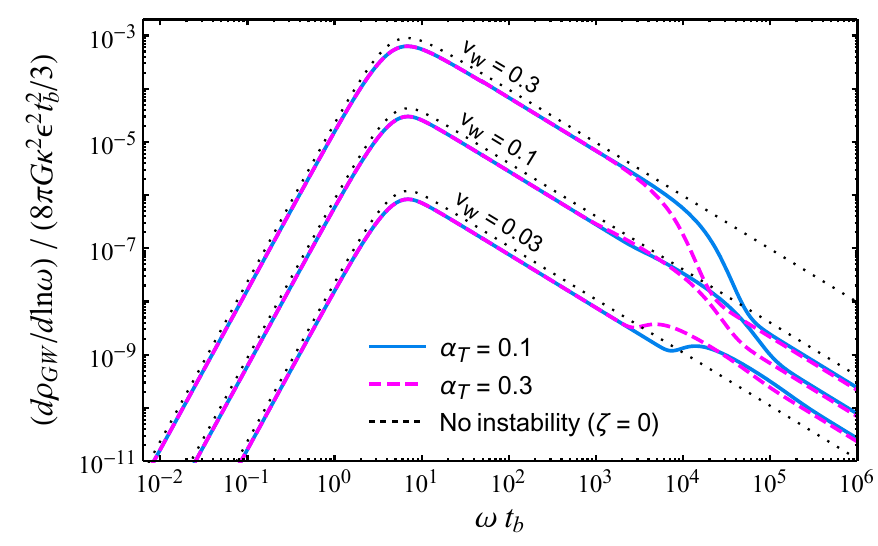}%
\caption{The GW spectrum for $Ht_b=0.05$. 
	For each value of $v_w$, the value of $a$ corresponds to $\langle(\nabla\zeta)^{2}\rangle=0.1$ and $v_{c}=2v_{w}$.%
\label{fig:Delta}}
\end{figure}

Figure \ref{fig:Delta} shows the total GW spectrum for different
values of the parameters $v_{w}$ and $\alpha_{T}$. 
The limiting case of $a=0$ is also shown. This corresponds to the bubble-collision mechanism without instability.
For $\omega\lesssim k_{*}$, the lowest-order contributions (\ref{Pi00sfinal})-(\ref{Pi00dfinal})
reproduce the bubble-collision result, 
while the corrections (\ref{Pi2002}) cause a global suppression.
For $\omega\gtrsim k_{*}$, these contributions decrease abruptly. 
The contribution (\ref{Pi11final}) is only relevant at $\omega\gtrsim k_*$.
It has a peak at $\omega\sim k_{*}$, which is more pronounced for small wall velocities.
For $k\gg k_*$, this contribution decreases as $\omega^{-1}$.

\begin{figure}[h]
	\centering
	\includegraphics[width=0.9\columnwidth]{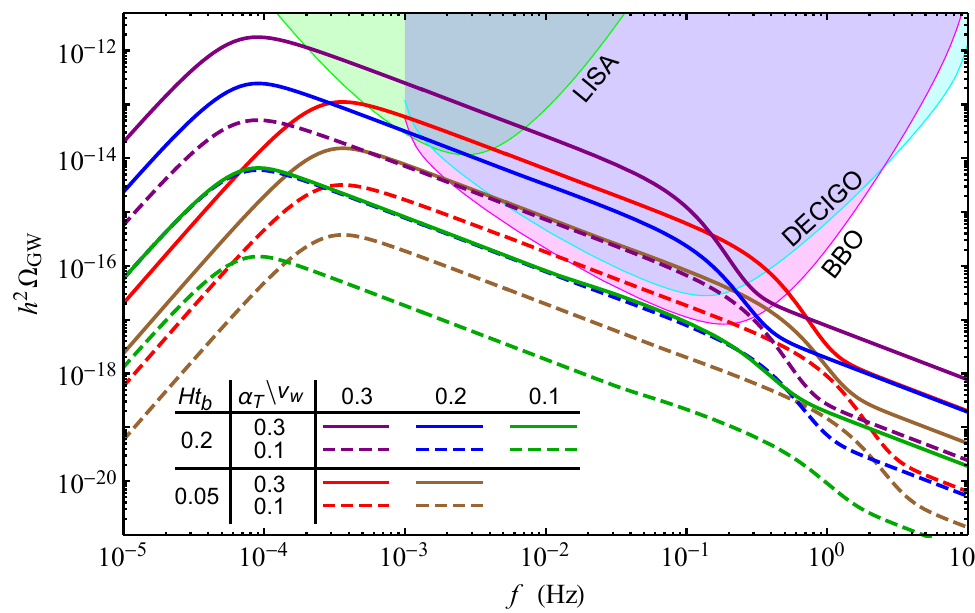}%
	\caption{The GW spectrum today for $v_{c}=2v_{w}$, and $\langle(\nabla\zeta)^{2}\rangle=0.1$.
		We include sensitivity curves for LISA \cite{a+17laser},
		BBO \cite{cc06detecting} and DECIGO \cite{k+11japanese},
		constructed from power-law integrated curves \cite{tr13sensitivity} using signal-to-noise ratio 1 and observation time 10 years.
		\label{fig:spectra}}
\end{figure}

To obtain the GW spectrum today, we multiply 
the frequency by the ratio of the scale factor at the time of GW generation, $a_*$,
to its present value, $a_0$,
and we multiply the energy density $\rho_{GW}$ by $(a_*/a_0)^4$
(see \cite{Largo} for details).
The efficiency factor $\kappa$ is a function of $v_w$ and $\alpha_T$.
We use the fit from Ref.~\cite{ekn+10energy}.
In Fig.~\ref{fig:spectra}
we consider the quantity $	\Omega_{GW} = \rho_{c}^{-1}{d\rho_{GW}}/{d\log f}$,
where $f$ is the frequency and $\rho_{c}=3H_{0}^{2}/(8\pi G)$ is the present value of
the critical density.
The Hubble parameter is given by
$H_{0}=100h\mathrm{km}/\mathrm{s}/\mathrm{Mpc}$, with $h\simeq 0.7$ \cite{n+24review}.
The figure also shows the sensitivity curves for future GW experiments.

In summary, we calculated the gravitational radiation produced
by bubble walls and thin fluid shells where perturbations grow
due to hydrodynamic instability. 
Assuming that unstable modes cease growing once they become nonlinear,
we performed a perturbative calculation in powers of the wall perturbation $\zeta$.
Depending on the parameters, this instability produces a signal in the GW spectrum at a frequency that is three or four orders of magnitude higher than the bubble-collision peak frequency.
Today, this frequency is around $10^{-1}\,\mathrm{Hz}$
and may be within reach of future space-based detectors such as BBO or DECIGO.
Due to our approximations, we have considered a small value of $\langle(\nabla\zeta)^{2}\rangle$
as well as relatively small values of $v_w$ and $\alpha_T$.
Higher values of these quantities will produce a stronger signal at lower frequencies.

\begin{acknowledgments}
This work was supported by UNMdP grant 80020250500055MP.
\end{acknowledgments}

\bibliography{C:/Users/ariel/OneDrive/Biblio/Catalogos/papers,C:/Users/ariel/OneDrive/Biblio/Catalogos/libros-reviews}

\begin{thebibliography}{40}%
\makeatletter
\providecommand \@ifxundefined [1]{%
 \@ifx{#1\undefined}
}%
\providecommand \@ifnum [1]{%
 \ifnum #1\expandafter \@firstoftwo
 \else \expandafter \@secondoftwo
 \fi
}%
\providecommand \@ifx [1]{%
 \ifx #1\expandafter \@firstoftwo
 \else \expandafter \@secondoftwo
 \fi
}%
\providecommand \natexlab [1]{#1}%
\providecommand \enquote  [1]{``#1''}%
\providecommand \bibnamefont  [1]{#1}%
\providecommand \bibfnamefont [1]{#1}%
\providecommand \citenamefont [1]{#1}%
\providecommand \href@noop [0]{\@secondoftwo}%
\providecommand \href [0]{\begingroup \@sanitize@url \@href}%
\providecommand \@href[1]{\@@startlink{#1}\@@href}%
\providecommand \@@href[1]{\endgroup#1\@@endlink}%
\providecommand \@sanitize@url [0]{\catcode `\\12\catcode `\$12\catcode
  `\&12\catcode `\#12\catcode `\^12\catcode `\_12\catcode `\%12\relax}%
\providecommand \@@startlink[1]{}%
\providecommand \@@endlink[0]{}%
\providecommand \url  [0]{\begingroup\@sanitize@url \@url }%
\providecommand \@url [1]{\endgroup\@href {#1}{\urlprefix }}%
\providecommand \urlprefix  [0]{URL }%
\providecommand \Eprint [0]{\href }%
\providecommand \doibase [0]{https://doi.org/}%
\providecommand \selectlanguage [0]{\@gobble}%
\providecommand \bibinfo  [0]{\@secondoftwo}%
\providecommand \bibfield  [0]{\@secondoftwo}%
\providecommand \translation [1]{[#1]}%
\providecommand \BibitemOpen [0]{}%
\providecommand \bibitemStop [0]{}%
\providecommand \bibitemNoStop [0]{.\EOS\space}%
\providecommand \EOS [0]{\spacefactor3000\relax}%
\providecommand \BibitemShut  [1]{\csname bibitem#1\endcsname}%
\let\auto@bib@innerbib\@empty
%</preamble>
\bibitem [{\citenamefont {Quiros}(2007)}]{quiros07field}%
  \BibitemOpen
  \bibfield  {author} {\bibinfo {author} {\bibfnamefont {M.}~\bibnamefont
  {Quiros}},\ }\bibfield  {title} {\bibinfo {title} {{Field theory at finite
  temperature and phase transitions}},\ }\href@noop {} {\bibfield  {journal}
  {\bibinfo  {journal} {Acta Phys. Polon. B}\ }\textbf {\bibinfo {volume}
  {38}},\ \bibinfo {pages} {3661} (\bibinfo {year} {2007})}\BibitemShut
  {NoStop}%
\bibitem [{\citenamefont {Athron}\ \emph {et~al.}(2024)\citenamefont {Athron},
  \citenamefont {Bal\'azs}, \citenamefont {Fowlie}, \citenamefont {Morris},\
  and\ \citenamefont {Wu}}]{abf+24cosmological}%
  \BibitemOpen
  \bibfield  {author} {\bibinfo {author} {\bibfnamefont {P.}~\bibnamefont
  {Athron}}, \bibinfo {author} {\bibfnamefont {C.}~\bibnamefont {Bal\'azs}},
  \bibinfo {author} {\bibfnamefont {A.}~\bibnamefont {Fowlie}}, \bibinfo
  {author} {\bibfnamefont {L.}~\bibnamefont {Morris}},\ and\ \bibinfo {author}
  {\bibfnamefont {L.}~\bibnamefont {Wu}},\ }\bibfield  {title} {\bibinfo
  {title} {{Cosmological phase transitions: From perturbative particle physics
  to gravitational waves}},\ }\href
  {https://doi.org/10.1016/j.ppnp.2023.104094} {\bibfield  {journal} {\bibinfo
  {journal} {Prog. Part. Nucl. Phys.}\ }\textbf {\bibinfo {volume} {135}},\
  \bibinfo {pages} {104094} (\bibinfo {year} {2024})},\ \Eprint
  {https://arxiv.org/abs/2305.02357} {arXiv:2305.02357 [hep-ph]} \BibitemShut
  {NoStop}%
\bibitem [{\citenamefont {Hogan}(1986)}]{hogan86gravitational}%
  \BibitemOpen
  \bibfield  {author} {\bibinfo {author} {\bibfnamefont {C.~J.}\ \bibnamefont
  {Hogan}},\ }\bibfield  {title} {\bibinfo {title} {{Gravitational radiation
  from cosmological phase transitions}},\ }\href@noop {} {\bibfield  {journal}
  {\bibinfo  {journal} {Mon. Not. Roy. Astron. Soc.}\ }\textbf {\bibinfo
  {volume} {218}},\ \bibinfo {pages} {629} (\bibinfo {year}
  {1986})}\BibitemShut {NoStop}%
\bibitem [{\citenamefont {Turner}\ and\ \citenamefont
  {Wilczek}(1990)}]{tw90relic}%
  \BibitemOpen
  \bibfield  {author} {\bibinfo {author} {\bibfnamefont {M.~S.}\ \bibnamefont
  {Turner}}\ and\ \bibinfo {author} {\bibfnamefont {F.}~\bibnamefont
  {Wilczek}},\ }\bibfield  {title} {\bibinfo {title} {{Relic gravitational
  waves and extended inflation}},\ }\href
  {https://doi.org/10.1103/PhysRevLett.65.3080} {\bibfield  {journal} {\bibinfo
   {journal} {Phys. Rev. Lett.}\ }\textbf {\bibinfo {volume} {65}},\ \bibinfo
  {pages} {3080} (\bibinfo {year} {1990})}\BibitemShut {NoStop}%
\bibitem [{\citenamefont {Amaro-Seoane}\ \emph {et~al.}(2017)\citenamefont
  {Amaro-Seoane} \emph {et~al.}}]{a+17laser}%
  \BibitemOpen
  \bibfield  {author} {\bibinfo {author} {\bibfnamefont {P.}~\bibnamefont
  {Amaro-Seoane}} \emph {et~al.} (\bibinfo {collaboration} {LISA}),\
  }\href@noop {} {\bibinfo {title} {{Laser Interferometer Space Antenna}}}
  (\bibinfo {year} {2017}),\ \Eprint {https://arxiv.org/abs/1702.00786}
  {arXiv:1702.00786 [astro-ph.IM]} \BibitemShut {NoStop}%
\bibitem [{\citenamefont {Corbin}\ and\ \citenamefont
  {Cornish}(2006)}]{cc06detecting}%
  \BibitemOpen
  \bibfield  {author} {\bibinfo {author} {\bibfnamefont {V.}~\bibnamefont
  {Corbin}}\ and\ \bibinfo {author} {\bibfnamefont {N.~J.}\ \bibnamefont
  {Cornish}},\ }\bibfield  {title} {\bibinfo {title} {Detecting the cosmic
  gravitational wave background with the big bang observer},\ }\href
  {https://doi.org/doi:10.1088/0264-9381/23/7/014} {\bibfield  {journal}
  {\bibinfo  {journal} {Classical and Quantum Gravity}\ }\textbf {\bibinfo
  {volume} {23}},\ \bibinfo {pages} {2435} (\bibinfo {year}
  {2006})}\BibitemShut {NoStop}%
\bibitem [{\citenamefont {Kawamura}\ \emph {et~al.}(2011)\citenamefont
  {Kawamura} \emph {et~al.}}]{k+11japanese}%
  \BibitemOpen
  \bibfield  {author} {\bibinfo {author} {\bibfnamefont {S.}~\bibnamefont
  {Kawamura}} \emph {et~al.},\ }\bibfield  {title} {\bibinfo {title} {{The
  Japanese space gravitational wave antenna: DECIGO}},\ }\href
  {https://doi.org/10.1088/0264-9381/28/9/094011} {\bibfield  {journal}
  {\bibinfo  {journal} {Class. Quant. Grav.}\ }\textbf {\bibinfo {volume}
  {28}},\ \bibinfo {pages} {094011} (\bibinfo {year} {2011})}\BibitemShut
  {NoStop}%
\bibitem [{\citenamefont {Kamionkowski}\ \emph {et~al.}(1994)\citenamefont
  {Kamionkowski}, \citenamefont {Kosowsky},\ and\ \citenamefont
  {Turner}}]{kkt94gravitational}%
  \BibitemOpen
  \bibfield  {author} {\bibinfo {author} {\bibfnamefont {M.}~\bibnamefont
  {Kamionkowski}}, \bibinfo {author} {\bibfnamefont {A.}~\bibnamefont
  {Kosowsky}},\ and\ \bibinfo {author} {\bibfnamefont {M.~S.}\ \bibnamefont
  {Turner}},\ }\bibfield  {title} {\bibinfo {title} {{Gravitational radiation
  from first order phase transitions}},\ }\href
  {https://doi.org/10.1103/PhysRevD.49.2837} {\bibfield  {journal} {\bibinfo
  {journal} {Phys. Rev. D}\ }\textbf {\bibinfo {volume} {49}},\ \bibinfo
  {pages} {2837} (\bibinfo {year} {1994})},\ \Eprint
  {https://arxiv.org/abs/astro-ph/9310044} {arXiv:astro-ph/9310044}
  \BibitemShut {NoStop}%
\bibitem [{\citenamefont {Apreda}\ \emph {et~al.}(2001)\citenamefont {Apreda},
  \citenamefont {Maggiore}, \citenamefont {Nicolis},\ and\ \citenamefont
  {Riotto}}]{amn+01supersymmetric}%
  \BibitemOpen
  \bibfield  {author} {\bibinfo {author} {\bibfnamefont {R.}~\bibnamefont
  {Apreda}}, \bibinfo {author} {\bibfnamefont {M.}~\bibnamefont {Maggiore}},
  \bibinfo {author} {\bibfnamefont {A.}~\bibnamefont {Nicolis}},\ and\ \bibinfo
  {author} {\bibfnamefont {A.}~\bibnamefont {Riotto}},\ }\bibfield  {title}
  {\bibinfo {title} {{Supersymmetric phase transitions and gravitational waves
  at LISA}},\ }\href {https://doi.org/10.1088/0264-9381/18/23/101} {\bibfield
  {journal} {\bibinfo  {journal} {Class. Quant. Grav.}\ }\textbf {\bibinfo
  {volume} {18}},\ \bibinfo {pages} {L155} (\bibinfo {year} {2001})},\ \Eprint
  {https://arxiv.org/abs/hep-ph/0102140} {arXiv:hep-ph/0102140} \BibitemShut
  {NoStop}%
\bibitem [{\citenamefont {Kajantie}\ \emph {et~al.}(1996)\citenamefont
  {Kajantie}, \citenamefont {Laine}, \citenamefont {Rummukainen},\ and\
  \citenamefont {Shaposhnikov}}]{klr+96is}%
  \BibitemOpen
  \bibfield  {author} {\bibinfo {author} {\bibfnamefont {K.}~\bibnamefont
  {Kajantie}}, \bibinfo {author} {\bibfnamefont {M.}~\bibnamefont {Laine}},
  \bibinfo {author} {\bibfnamefont {K.}~\bibnamefont {Rummukainen}},\ and\
  \bibinfo {author} {\bibfnamefont {M.~E.}\ \bibnamefont {Shaposhnikov}},\
  }\bibfield  {title} {\bibinfo {title} {{Is there a~ hot electroweak phase
  transition at $m_H \gtrsim m_W$?}},\ }\href
  {https://doi.org/10.1103/PhysRevLett.77.2887} {\bibfield  {journal} {\bibinfo
   {journal} {Phys. Rev. Lett.}\ }\textbf {\bibinfo {volume} {77}},\ \bibinfo
  {pages} {2887} (\bibinfo {year} {1996})},\ \Eprint
  {https://arxiv.org/abs/hep-ph/9605288} {arXiv:hep-ph/9605288} \BibitemShut
  {NoStop}%
\bibitem [{\citenamefont {Carena}\ \emph {et~al.}(1996)\citenamefont {Carena},
  \citenamefont {Quiros},\ and\ \citenamefont {Wagner}}]{cqw96opening}%
  \BibitemOpen
  \bibfield  {author} {\bibinfo {author} {\bibfnamefont {M.}~\bibnamefont
  {Carena}}, \bibinfo {author} {\bibfnamefont {M.}~\bibnamefont {Quiros}},\
  and\ \bibinfo {author} {\bibfnamefont {C.~E.~M.}\ \bibnamefont {Wagner}},\
  }\bibfield  {title} {\bibinfo {title} {{Opening the window for electroweak
  baryogenesis}},\ }\href {https://doi.org/10.1016/0370-2693(96)00475-3}
  {\bibfield  {journal} {\bibinfo  {journal} {Phys. Lett. B}\ }\textbf
  {\bibinfo {volume} {380}},\ \bibinfo {pages} {81} (\bibinfo {year} {1996})},\
  \Eprint {https://arxiv.org/abs/hep-ph/9603420} {arXiv:hep-ph/9603420}
  \BibitemShut {NoStop}%
\bibitem [{\citenamefont {Grojean}\ \emph {et~al.}(2005)\citenamefont
  {Grojean}, \citenamefont {Servant},\ and\ \citenamefont
  {Wells}}]{gsw05first}%
  \BibitemOpen
  \bibfield  {author} {\bibinfo {author} {\bibfnamefont {C.}~\bibnamefont
  {Grojean}}, \bibinfo {author} {\bibfnamefont {G.}~\bibnamefont {Servant}},\
  and\ \bibinfo {author} {\bibfnamefont {J.~D.}\ \bibnamefont {Wells}},\
  }\bibfield  {title} {\bibinfo {title} {{First-order electroweak phase
  transition in the standard model with a low cutoff}},\ }\href
  {https://doi.org/10.1103/PhysRevD.71.036001} {\bibfield  {journal} {\bibinfo
  {journal} {Phys. Rev. D}\ }\textbf {\bibinfo {volume} {71}},\ \bibinfo
  {pages} {036001} (\bibinfo {year} {2005})},\ \Eprint
  {https://arxiv.org/abs/hep-ph/0407019} {arXiv:hep-ph/0407019} \BibitemShut
  {NoStop}%
\bibitem [{\citenamefont {Espinosa}\ \emph {et~al.}(2008)\citenamefont
  {Espinosa}, \citenamefont {Konstandin}, \citenamefont {No},\ and\
  \citenamefont {Quiros}}]{ekn+08some}%
  \BibitemOpen
  \bibfield  {author} {\bibinfo {author} {\bibfnamefont {J.~R.}\ \bibnamefont
  {Espinosa}}, \bibinfo {author} {\bibfnamefont {T.}~\bibnamefont
  {Konstandin}}, \bibinfo {author} {\bibfnamefont {J.~M.}\ \bibnamefont {No}},\
  and\ \bibinfo {author} {\bibfnamefont {M.}~\bibnamefont {Quiros}},\
  }\bibfield  {title} {\bibinfo {title} {{Some Cosmological Implications of
  Hidden Sectors}},\ }\href {https://doi.org/10.1103/PhysRevD.78.123528}
  {\bibfield  {journal} {\bibinfo  {journal} {Phys. Rev. D}\ }\textbf {\bibinfo
  {volume} {78}},\ \bibinfo {pages} {123528} (\bibinfo {year} {2008})},\
  \Eprint {https://arxiv.org/abs/0809.3215} {arXiv:0809.3215 [hep-ph]}
  \BibitemShut {NoStop}%
\bibitem [{\citenamefont {Huber}\ and\ \citenamefont
  {Konstandin}(2008)}]{hk08production}%
  \BibitemOpen
  \bibfield  {author} {\bibinfo {author} {\bibfnamefont {S.~J.}\ \bibnamefont
  {Huber}}\ and\ \bibinfo {author} {\bibfnamefont {T.}~\bibnamefont
  {Konstandin}},\ }\bibfield  {title} {\bibinfo {title} {{Production of
  gravitational waves in the nMSSM}},\ }\href
  {https://doi.org/10.1088/1475-7516/2008/05/017} {\bibfield  {journal}
  {\bibinfo  {journal} {JCAP}\ }\textbf {\bibinfo {volume} {2008}}\bibfield
  {number} {\bibinfo  {number} { (05)},\ \bibinfo {pages} {017}},\ }\Eprint
  {https://arxiv.org/abs/0709.2091} {arXiv:0709.2091 [hep-ph]} \BibitemShut
  {NoStop}%
\bibitem [{\citenamefont {Dorsch}\ \emph {et~al.}(2017)\citenamefont {Dorsch},
  \citenamefont {Huber}, \citenamefont {Konstandin},\ and\ \citenamefont
  {No}}]{dhk+17second}%
  \BibitemOpen
  \bibfield  {author} {\bibinfo {author} {\bibfnamefont {G.~C.}\ \bibnamefont
  {Dorsch}}, \bibinfo {author} {\bibfnamefont {S.~J.}\ \bibnamefont {Huber}},
  \bibinfo {author} {\bibfnamefont {T.}~\bibnamefont {Konstandin}},\ and\
  \bibinfo {author} {\bibfnamefont {J.~M.}\ \bibnamefont {No}},\ }\bibfield
  {title} {\bibinfo {title} {{A Second Higgs Doublet in the Early Universe:
  Baryogenesis and Gravitational Waves}},\ }\href
  {https://doi.org/10.1088/1475-7516/2017/05/052} {\bibfield  {journal}
  {\bibinfo  {journal} {JCAP}\ }\textbf {\bibinfo {volume} {2017}}\bibfield
  {number} {\bibinfo  {number} { (05)},\ \bibinfo {pages} {052}},\ }\Eprint
  {https://arxiv.org/abs/1611.05874} {arXiv:1611.05874 [hep-ph]} \BibitemShut
  {NoStop}%
\bibitem [{\citenamefont {Meg\'\i{}as}\ \emph {et~al.}(2018)\citenamefont
  {Meg\'\i{}as}, \citenamefont {Nardini},\ and\ \citenamefont
  {Quir\'os}}]{mnq18cosmological}%
  \BibitemOpen
  \bibfield  {author} {\bibinfo {author} {\bibfnamefont {E.}~\bibnamefont
  {Meg\'\i{}as}}, \bibinfo {author} {\bibfnamefont {G.}~\bibnamefont
  {Nardini}},\ and\ \bibinfo {author} {\bibfnamefont {M.}~\bibnamefont
  {Quir\'os}},\ }\bibfield  {title} {\bibinfo {title} {{Cosmological Phase
  Transitions in Warped Space: Gravitational Waves and Collider Signatures}},\
  }\href {https://doi.org/10.1007/JHEP09(2018)095} {\bibfield  {journal}
  {\bibinfo  {journal} {JHEP}\ }\textbf {\bibinfo {volume} {2018}}\bibfield
  {number} {\bibinfo  {number} { (09)},\ \bibinfo {pages} {095}},\ }\Eprint
  {https://arxiv.org/abs/1806.04877} {arXiv:1806.04877 [hep-ph]} \BibitemShut
  {NoStop}%
\bibitem [{\citenamefont {Kosowsky}\ \emph {et~al.}(1992)\citenamefont
  {Kosowsky}, \citenamefont {Turner},\ and\ \citenamefont
  {Watkins}}]{ktw92gravitational}%
  \BibitemOpen
  \bibfield  {author} {\bibinfo {author} {\bibfnamefont {A.}~\bibnamefont
  {Kosowsky}}, \bibinfo {author} {\bibfnamefont {M.~S.}\ \bibnamefont
  {Turner}},\ and\ \bibinfo {author} {\bibfnamefont {R.}~\bibnamefont
  {Watkins}},\ }\bibfield  {title} {\bibinfo {title} {{Gravitational waves from
  first order cosmological phase transitions}},\ }\href
  {https://doi.org/10.1103/PhysRevLett.69.2026} {\bibfield  {journal} {\bibinfo
   {journal} {Phys. Rev. Lett.}\ }\textbf {\bibinfo {volume} {69}},\ \bibinfo
  {pages} {2026} (\bibinfo {year} {1992})}\BibitemShut {NoStop}%
\bibitem [{\citenamefont {Kosowsky}\ and\ \citenamefont
  {Turner}(1993)}]{kt93gravitational}%
  \BibitemOpen
  \bibfield  {author} {\bibinfo {author} {\bibfnamefont {A.}~\bibnamefont
  {Kosowsky}}\ and\ \bibinfo {author} {\bibfnamefont {M.~S.}\ \bibnamefont
  {Turner}},\ }\bibfield  {title} {\bibinfo {title} {{Gravitational radiation
  from colliding vacuum bubbles: envelope approximation to many bubble
  collisions}},\ }\href {https://doi.org/10.1103/PhysRevD.47.4372} {\bibfield
  {journal} {\bibinfo  {journal} {Phys. Rev.}\ }\textbf {\bibinfo {volume}
  {D47}},\ \bibinfo {pages} {4372} (\bibinfo {year} {1993})},\ \Eprint
  {https://arxiv.org/abs/astro-ph/9211004} {arXiv:astro-ph/9211004 [astro-ph]}
  \BibitemShut {NoStop}%
%%CITATION = ASTRO-PH/9211004;%%
\bibitem [{\citenamefont {Dolgov}\ \emph {et~al.}(2002)\citenamefont {Dolgov},
  \citenamefont {Grasso},\ and\ \citenamefont {Nicolis}}]{dgn02relic}%
  \BibitemOpen
  \bibfield  {author} {\bibinfo {author} {\bibfnamefont {A.~D.}\ \bibnamefont
  {Dolgov}}, \bibinfo {author} {\bibfnamefont {D.}~\bibnamefont {Grasso}},\
  and\ \bibinfo {author} {\bibfnamefont {A.}~\bibnamefont {Nicolis}},\
  }\bibfield  {title} {\bibinfo {title} {{Relic backgrounds of gravitational
  waves from cosmic turbulence}},\ }\href
  {https://doi.org/10.1103/PhysRevD.66.103505} {\bibfield  {journal} {\bibinfo
  {journal} {Phys. Rev. D}\ }\textbf {\bibinfo {volume} {66}},\ \bibinfo
  {pages} {103505} (\bibinfo {year} {2002})},\ \Eprint
  {https://arxiv.org/abs/astro-ph/0206461} {arXiv:astro-ph/0206461}
  \BibitemShut {NoStop}%
\bibitem [{\citenamefont {Kosowsky}\ \emph {et~al.}(2002)\citenamefont
  {Kosowsky}, \citenamefont {Mack},\ and\ \citenamefont
  {Kahniashvili}}]{kmk02gravitational}%
  \BibitemOpen
  \bibfield  {author} {\bibinfo {author} {\bibfnamefont {A.}~\bibnamefont
  {Kosowsky}}, \bibinfo {author} {\bibfnamefont {A.}~\bibnamefont {Mack}},\
  and\ \bibinfo {author} {\bibfnamefont {T.}~\bibnamefont {Kahniashvili}},\
  }\bibfield  {title} {\bibinfo {title} {{Gravitational radiation from
  cosmological turbulence}},\ }\href
  {https://doi.org/10.1103/PhysRevD.66.024030} {\bibfield  {journal} {\bibinfo
  {journal} {Phys. Rev. D}\ }\textbf {\bibinfo {volume} {66}},\ \bibinfo
  {pages} {024030} (\bibinfo {year} {2002})},\ \Eprint
  {https://arxiv.org/abs/astro-ph/0111483} {arXiv:astro-ph/0111483}
  \BibitemShut {NoStop}%
\bibitem [{\citenamefont {Caprini}\ and\ \citenamefont
  {Durrer}(2006)}]{cd06gravitational}%
  \BibitemOpen
  \bibfield  {author} {\bibinfo {author} {\bibfnamefont {C.}~\bibnamefont
  {Caprini}}\ and\ \bibinfo {author} {\bibfnamefont {R.}~\bibnamefont
  {Durrer}},\ }\bibfield  {title} {\bibinfo {title} {{Gravitational waves from
  stochastic relativistic sources: Primordial turbulence and magnetic
  fields}},\ }\href {https://doi.org/10.1103/PhysRevD.74.063521} {\bibfield
  {journal} {\bibinfo  {journal} {Phys. Rev. D}\ }\textbf {\bibinfo {volume}
  {74}},\ \bibinfo {pages} {063521} (\bibinfo {year} {2006})},\ \Eprint
  {https://arxiv.org/abs/astro-ph/0603476} {arXiv:astro-ph/0603476}
  \BibitemShut {NoStop}%
\bibitem [{\citenamefont {Hindmarsh}\ \emph {et~al.}(2014)\citenamefont
  {Hindmarsh}, \citenamefont {Huber}, \citenamefont {Rummukainen},\ and\
  \citenamefont {Weir}}]{hhr+14gravitational}%
  \BibitemOpen
  \bibfield  {author} {\bibinfo {author} {\bibfnamefont {M.}~\bibnamefont
  {Hindmarsh}}, \bibinfo {author} {\bibfnamefont {S.~J.}\ \bibnamefont
  {Huber}}, \bibinfo {author} {\bibfnamefont {K.}~\bibnamefont {Rummukainen}},\
  and\ \bibinfo {author} {\bibfnamefont {D.~J.}\ \bibnamefont {Weir}},\
  }\bibfield  {title} {\bibinfo {title} {{Gravitational waves from the sound of
  a first order phase transition}},\ }\href
  {https://doi.org/10.1103/PhysRevLett.112.041301} {\bibfield  {journal}
  {\bibinfo  {journal} {Phys. Rev. Lett.}\ }\textbf {\bibinfo {volume} {112}},\
  \bibinfo {pages} {041301} (\bibinfo {year} {2014})},\ \Eprint
  {https://arxiv.org/abs/1304.2433} {arXiv:1304.2433 [hep-ph]} \BibitemShut
  {NoStop}%
\bibitem [{\citenamefont {Hindmarsh}(2018)}]{hindmarsh18sound}%
  \BibitemOpen
  \bibfield  {author} {\bibinfo {author} {\bibfnamefont {M.}~\bibnamefont
  {Hindmarsh}},\ }\bibfield  {title} {\bibinfo {title} {{Sound shell model for
  acoustic gravitational wave production at a first-order phase transition in
  the early Universe}},\ }\href
  {https://doi.org/10.1103/PhysRevLett.120.071301} {\bibfield  {journal}
  {\bibinfo  {journal} {Phys. Rev. Lett.}\ }\textbf {\bibinfo {volume} {120}},\
  \bibinfo {pages} {071301} (\bibinfo {year} {2018})},\ \Eprint
  {https://arxiv.org/abs/1608.04735} {arXiv:1608.04735 [astro-ph.CO]}
  \BibitemShut {NoStop}%
\bibitem [{\citenamefont {Landau}\ and\ \citenamefont
  {Lifshitz}(1987)}]{LandauLifshitz87Fluid}%
  \BibitemOpen
  \bibfield  {author} {\bibinfo {author} {\bibfnamefont {L.~D.}\ \bibnamefont
  {Landau}}\ and\ \bibinfo {author} {\bibfnamefont {E.~M.}\ \bibnamefont
  {Lifshitz}},\ }\href@noop {} {\emph {\bibinfo {title} {Fluid Mechanics}}},\
  \bibinfo {edition} {2nd}\ ed.\ (\bibinfo  {publisher} {Pergamon Press},\
  \bibinfo {address} {London},\ \bibinfo {year} {1987})\BibitemShut {NoStop}%
\bibitem [{\citenamefont {Freese}\ and\ \citenamefont
  {Adams}(1990)}]{fa90hadron}%
  \BibitemOpen
  \bibfield  {author} {\bibinfo {author} {\bibfnamefont {K.}~\bibnamefont
  {Freese}}\ and\ \bibinfo {author} {\bibfnamefont {F.~C.}\ \bibnamefont
  {Adams}},\ }\bibfield  {title} {\bibinfo {title} {{Hadron Bubble Evolution
  Into the Quark Sea}},\ }\href {https://doi.org/10.1103/PhysRevD.41.2449}
  {\bibfield  {journal} {\bibinfo  {journal} {Phys. Rev. D}\ }\textbf {\bibinfo
  {volume} {41}},\ \bibinfo {pages} {2449} (\bibinfo {year}
  {1990})}\BibitemShut {NoStop}%
\bibitem [{\citenamefont {Link}(1992)}]{link92deflagration}%
  \BibitemOpen
  \bibfield  {author} {\bibinfo {author} {\bibfnamefont {B.}~\bibnamefont
  {Link}},\ }\bibfield  {title} {\bibinfo {title} {{Deflagration instability in
  the quark - hadron phase transition}},\ }\href
  {https://doi.org/10.1103/PhysRevLett.68.2425} {\bibfield  {journal} {\bibinfo
   {journal} {Phys. Rev. Lett.}\ }\textbf {\bibinfo {volume} {68}},\ \bibinfo
  {pages} {2425} (\bibinfo {year} {1992})}\BibitemShut {NoStop}%
%%CITATION = PRLTA,68,2425;%%
\bibitem [{\citenamefont {Kamionkowski}\ and\ \citenamefont
  {Freese}(1992)}]{kf92instability}%
  \BibitemOpen
  \bibfield  {author} {\bibinfo {author} {\bibfnamefont {M.}~\bibnamefont
  {Kamionkowski}}\ and\ \bibinfo {author} {\bibfnamefont {K.}~\bibnamefont
  {Freese}},\ }\bibfield  {title} {\bibinfo {title} {{Instability and
  subsequent evolution of electroweak bubbles}},\ }\href
  {https://doi.org/10.1103/PhysRevLett.69.2743} {\bibfield  {journal} {\bibinfo
   {journal} {Phys. Rev. Lett.}\ }\textbf {\bibinfo {volume} {69}},\ \bibinfo
  {pages} {2743} (\bibinfo {year} {1992})},\ \Eprint
  {https://arxiv.org/abs/hep-ph/9208202} {arXiv:hep-ph/9208202} \BibitemShut
  {NoStop}%
\bibitem [{\citenamefont {Huet}\ \emph {et~al.}(1993)\citenamefont {Huet},
  \citenamefont {Kajantie}, \citenamefont {Leigh}, \citenamefont {Liu},\ and\
  \citenamefont {McLerran}}]{hkl+93hydrodynamic}%
  \BibitemOpen
  \bibfield  {author} {\bibinfo {author} {\bibfnamefont {P.~Y.}\ \bibnamefont
  {Huet}}, \bibinfo {author} {\bibfnamefont {K.}~\bibnamefont {Kajantie}},
  \bibinfo {author} {\bibfnamefont {R.~G.}\ \bibnamefont {Leigh}}, \bibinfo
  {author} {\bibfnamefont {B.-H.}\ \bibnamefont {Liu}},\ and\ \bibinfo {author}
  {\bibfnamefont {L.~D.}\ \bibnamefont {McLerran}},\ }\bibfield  {title}
  {\bibinfo {title} {{Hydrodynamic stability analysis of burning bubbles in
  electroweak theory and in QCD}},\ }\href
  {https://doi.org/10.1103/PhysRevD.48.2477} {\bibfield  {journal} {\bibinfo
  {journal} {Phys. Rev.}\ }\textbf {\bibinfo {volume} {D48}},\ \bibinfo {pages}
  {2477} (\bibinfo {year} {1993})},\ \Eprint
  {https://arxiv.org/abs/hep-ph/9212224} {arXiv:hep-ph/9212224 [hep-ph]}
  \BibitemShut {NoStop}%
%%CITATION = HEP-PH/9212224;%%
\bibitem [{\citenamefont {Megevand}\ and\ \citenamefont
  {Membiela}(2014)}]{mm14stability}%
  \BibitemOpen
  \bibfield  {author} {\bibinfo {author} {\bibfnamefont {A.}~\bibnamefont
  {Megevand}}\ and\ \bibinfo {author} {\bibfnamefont {F.~A.}\ \bibnamefont
  {Membiela}},\ }\bibfield  {title} {\bibinfo {title} {Stability of
  cosmological deflagration fronts},\ }\href
  {https://doi.org/10.1103/PhysRevD.89.103507} {\bibfield  {journal} {\bibinfo
  {journal} {Phys. Rev.}\ }\textbf {\bibinfo {volume} {D89}},\ \bibinfo {pages}
  {103507} (\bibinfo {year} {2014})},\ \Eprint
  {https://arxiv.org/abs/1311.2453} {arXiv:1311.2453 [astro-ph.CO]}
  \BibitemShut {NoStop}%
%%CITATION = ARXIV:1311.2453;%%
\bibitem [{\citenamefont {Megevand}\ and\ \citenamefont
  {Membiela}(2026)}]{Largo}%
  \BibitemOpen
  \bibfield  {author} {\bibinfo {author} {\bibfnamefont {A.}~\bibnamefont
  {Megevand}}\ and\ \bibinfo {author} {\bibfnamefont {F.~A.}\ \bibnamefont
  {Membiela}},\ }\bibfield  {title} {\bibinfo {title} {Gravitational waves from
  corrugated bubble walls}} (\bibinfo {year} {2026}),\ \bibinfo {note} {in
  preparation}\BibitemShut {NoStop}%
\bibitem [{\citenamefont {Megevand}\ and\ \citenamefont
  {Membiela}(2021{\natexlab{a}})}]{mm21model}%
  \BibitemOpen
  \bibfield  {author} {\bibinfo {author} {\bibfnamefont {A.}~\bibnamefont
  {Megevand}}\ and\ \bibinfo {author} {\bibfnamefont {F.~A.}\ \bibnamefont
  {Membiela}},\ }\bibfield  {title} {\bibinfo {title} {{Model-independent
  features of gravitational waves from bubble collisions}},\ }\href
  {https://doi.org/10.1103/PhysRevD.104.123532} {\bibfield  {journal} {\bibinfo
   {journal} {Phys. Rev. D}\ }\textbf {\bibinfo {volume} {104}},\ \bibinfo
  {pages} {123532} (\bibinfo {year} {2021}{\natexlab{a}})},\ \Eprint
  {https://arxiv.org/abs/2108.07034} {arXiv:2108.07034 [astro-ph.CO]}
  \BibitemShut {NoStop}%
\bibitem [{\citenamefont {Sigl}\ \emph {et~al.}(1997)\citenamefont {Sigl},
  \citenamefont {Olinto},\ and\ \citenamefont {Jedamzik}}]{soj97primordial}%
  \BibitemOpen
  \bibfield  {author} {\bibinfo {author} {\bibfnamefont {G.}~\bibnamefont
  {Sigl}}, \bibinfo {author} {\bibfnamefont {A.~V.}\ \bibnamefont {Olinto}},\
  and\ \bibinfo {author} {\bibfnamefont {K.}~\bibnamefont {Jedamzik}},\
  }\bibfield  {title} {\bibinfo {title} {{Primordial magnetic fields from
  cosmological first order phase transitions}},\ }\href
  {https://doi.org/10.1103/PhysRevD.55.4582} {\bibfield  {journal} {\bibinfo
  {journal} {Phys. Rev.}\ }\textbf {\bibinfo {volume} {D55}},\ \bibinfo {pages}
  {4582} (\bibinfo {year} {1997})},\ \Eprint
  {https://arxiv.org/abs/astro-ph/9610201} {arXiv:astro-ph/9610201 [astro-ph]}
  \BibitemShut {NoStop}%
%%CITATION = ASTRO-PH/9610201;%%
\bibitem [{\citenamefont {M\'egevand}\ and\ \citenamefont
  {Membiela}(2023)}]{mm23thin}%
  \BibitemOpen
  \bibfield  {author} {\bibinfo {author} {\bibfnamefont {A.}~\bibnamefont
  {M\'egevand}}\ and\ \bibinfo {author} {\bibfnamefont {F.~A.}\ \bibnamefont
  {Membiela}},\ }\bibfield  {title} {\bibinfo {title} {{Thin and thick bubble
  walls. Part I. Vacuum phase transitions}},\ }\href
  {https://doi.org/10.1088/1475-7516/2023/06/007} {\bibfield  {journal}
  {\bibinfo  {journal} {JCAP}\ }\textbf {\bibinfo {volume} {2023}}\bibfield
  {number} {\bibinfo  {number} { (06)},\ \bibinfo {pages} {007}},\ }\Eprint
  {https://arxiv.org/abs/2302.13349} {arXiv:2302.13349 [gr-qc]} \BibitemShut
  {NoStop}%
\bibitem [{\citenamefont {M{\'e}gevand}\ and\ \citenamefont
  {Membiela}(2025)}]{mm25thin}%
  \BibitemOpen
  \bibfield  {author} {\bibinfo {author} {\bibfnamefont {A.}~\bibnamefont
  {M{\'e}gevand}}\ and\ \bibinfo {author} {\bibfnamefont {F.~A.}\ \bibnamefont
  {Membiela}},\ }\bibfield  {title} {\bibinfo {title} {{Thin and thick bubble
  walls. Part III. Wall energy}},\ }\href
  {https://doi.org/10.1088/1475-7516/2025/05/076} {\bibfield  {journal}
  {\bibinfo  {journal} {JCAP}\ }\textbf {\bibinfo {volume} {2025}}\bibfield
  {number} {\bibinfo  {number} { (05)},\ \bibinfo {pages} {076}},\ }\Eprint
  {https://arxiv.org/abs/2501.05612} {arXiv:2501.05612 [gr-qc]} \BibitemShut
  {NoStop}%
\bibitem [{\citenamefont {Megevand}\ and\ \citenamefont
  {Membiela}(2021{\natexlab{b}})}]{mm21gravitational}%
  \BibitemOpen
  \bibfield  {author} {\bibinfo {author} {\bibfnamefont {A.}~\bibnamefont
  {Megevand}}\ and\ \bibinfo {author} {\bibfnamefont {F.~A.}\ \bibnamefont
  {Membiela}},\ }\bibfield  {title} {\bibinfo {title} {{Gravitational waves
  from bubble walls}},\ }\href {https://doi.org/10.1088/1475-7516/2021/10/073}
  {\bibfield  {journal} {\bibinfo  {journal} {JCAP}\ }\textbf {\bibinfo
  {volume} {2021}}\bibfield  {number} {\bibinfo  {number} { (10)},\ \bibinfo
  {pages} {073}},\ }\Eprint {https://arxiv.org/abs/2108.05510}
  {arXiv:2108.05510 [astro-ph.CO]} \BibitemShut {NoStop}%
\bibitem [{\citenamefont {Jinno}\ and\ \citenamefont
  {Takimoto}(2017)}]{jt17gravitational}%
  \BibitemOpen
  \bibfield  {author} {\bibinfo {author} {\bibfnamefont {R.}~\bibnamefont
  {Jinno}}\ and\ \bibinfo {author} {\bibfnamefont {M.}~\bibnamefont
  {Takimoto}},\ }\bibfield  {title} {\bibinfo {title} {{Gravitational waves
  from bubble collisions: An analytic derivation}},\ }\href
  {https://doi.org/10.1103/PhysRevD.95.024009} {\bibfield  {journal} {\bibinfo
  {journal} {Phys. Rev.}\ }\textbf {\bibinfo {volume} {D95}},\ \bibinfo {pages}
  {024009} (\bibinfo {year} {2017})},\ \Eprint
  {https://arxiv.org/abs/1605.01403} {arXiv:1605.01403 [astro-ph.CO]}
  \BibitemShut {NoStop}%
%%CITATION = ARXIV:1605.01403;%%
\bibitem [{\citenamefont {M\'egevand}\ and\ \citenamefont
  {Membiela}(2020)}]{mm20bubble}%
  \BibitemOpen
  \bibfield  {author} {\bibinfo {author} {\bibfnamefont {A.}~\bibnamefont
  {M\'egevand}}\ and\ \bibinfo {author} {\bibfnamefont {F.~A.}\ \bibnamefont
  {Membiela}},\ }\bibfield  {title} {\bibinfo {title} {{Bubble wall
  correlations in cosmological phase transitions}},\ }\href
  {https://doi.org/10.1103/PhysRevD.102.103514} {\bibfield  {journal} {\bibinfo
   {journal} {Phys. Rev. D}\ }\textbf {\bibinfo {volume} {102}},\ \bibinfo
  {pages} {103514} (\bibinfo {year} {2020})},\ \Eprint
  {https://arxiv.org/abs/2008.01873} {arXiv:2008.01873 [astro-ph.CO]}
  \BibitemShut {NoStop}%
\bibitem [{\citenamefont {Espinosa}\ \emph {et~al.}(2010)\citenamefont
  {Espinosa}, \citenamefont {Konstandin}, \citenamefont {No},\ and\
  \citenamefont {Servant}}]{ekn+10energy}%
  \BibitemOpen
  \bibfield  {author} {\bibinfo {author} {\bibfnamefont {J.~R.}\ \bibnamefont
  {Espinosa}}, \bibinfo {author} {\bibfnamefont {T.}~\bibnamefont
  {Konstandin}}, \bibinfo {author} {\bibfnamefont {J.~M.}\ \bibnamefont {No}},\
  and\ \bibinfo {author} {\bibfnamefont {G.}~\bibnamefont {Servant}},\
  }\bibfield  {title} {\bibinfo {title} {{Energy Budget of Cosmological
  First-order Phase Transitions}},\ }\href
  {https://doi.org/10.1088/1475-7516/2010/06/028} {\bibfield  {journal}
  {\bibinfo  {journal} {JCAP}\ }\textbf {\bibinfo {volume} {2010}}\bibfield
  {number} {\bibinfo  {number} { (06)},\ \bibinfo {pages} {028}},\ }\Eprint
  {https://arxiv.org/abs/1004.4187} {arXiv:1004.4187 [hep-ph]} \BibitemShut
  {NoStop}%
\bibitem [{\citenamefont {Navas}\ \emph {et~al.}(2024)\citenamefont {Navas}
  \emph {et~al.}}]{n+24review}%
  \BibitemOpen
  \bibfield  {author} {\bibinfo {author} {\bibfnamefont {S.}~\bibnamefont
  {Navas}} \emph {et~al.} (\bibinfo {collaboration} {Particle Data Group}),\
  }\bibfield  {title} {\bibinfo {title} {{Review of particle physics}},\ }\href
  {https://doi.org/10.1103/PhysRevD.110.030001} {\bibfield  {journal} {\bibinfo
   {journal} {Phys. Rev. D}\ }\textbf {\bibinfo {volume} {110}},\ \bibinfo
  {pages} {030001} (\bibinfo {year} {2024})}\BibitemShut {NoStop}%
\bibitem [{\citenamefont {Thrane}\ and\ \citenamefont
  {Romano}(2013)}]{tr13sensitivity}%
  \BibitemOpen
  \bibfield  {author} {\bibinfo {author} {\bibfnamefont {E.}~\bibnamefont
  {Thrane}}\ and\ \bibinfo {author} {\bibfnamefont {J.~D.}\ \bibnamefont
  {Romano}},\ }\bibfield  {title} {\bibinfo {title} {Sensitivity curves for
  searches for gravitational-wave backgrounds},\ }\bibfield  {journal}
  {\bibinfo  {journal} {Physical Review D}\ }\textbf {\bibinfo {volume} {88}},\
  \href {https://doi.org/10.1103/physrevd.88.124032}
  {10.1103/physrevd.88.124032} (\bibinfo {year} {2013})\BibitemShut {NoStop}%
\end{thebibliography}%

\end{document}